\documentclass[a4paper]{lipics-v2016}
\usepackage[utf8]{inputenc}
\usepackage{microtype}
\usepackage{listings}
\usepackage{srcltx}
\usepackage[ttscale=.875]{libertine}

\newcommand\YAMLcolonstyle{\color{red}\mdseries}
\newcommand\YAMLkeystyle{\color{black}\bfseries}
\newcommand\YAMLvaluestyle{\color{blue}\mdseries}

\makeatletter

\newcommand\language@yaml{yaml}

\expandafter\expandafter\expandafter\lstdefinelanguage
\expandafter{\language@yaml}
{
  keywords={true,false,null,y,n},
  keywordstyle=\color{darkgray}\bfseries,
  basicstyle=\YAMLkeystyle,                                 
  sensitive=false,
  comment=[l]{\#},
  morecomment=[s]{/*}{*/},
  commentstyle=\color{purple}\ttfamily,
  stringstyle=\YAMLvaluestyle\ttfamily,
  moredelim=[l][\color{orange}]{\&},
  moredelim=[l][\color{magenta}]{*},
  moredelim=**[il][\YAMLcolonstyle{:}\YAMLvaluestyle]{:},   
  morestring=[b]',
  morestring=[b]",
  literate =    {---}{{\ProcessThreeDashes}}3
                {>}{{\textcolor{red}\textgreater}}1     
                {|}{{\textcolor{red}\textbar}}1 
                {\ -\ }{{\mdseries\ -\ }}3,
}

\lst@AddToHook{EveryLine}{\ifx\lst@language\language@yaml\YAMLkeystyle\fi}
\makeatother

\newcommand\ProcessThreeDashes{\llap{\color{cyan}\mdseries-{-}-}}

\let\accentvec\vec

\let\vec\accentvec

\usepackage{amsmath}
\usepackage{amssymb}
\usepackage{amsfonts}
\usepackage{verbatim}
\usepackage[textsize=tiny]{todonotes}

\authorrunning{M. Aumüller, E. Bernhardsson, A. Faithfull}

\Copyright{ Martin Aumüller, Erik Bernhardsson, Alexander Faithfull} 



\usepackage{pgfplots}
\usepgfplotslibrary{groupplots,colorbrewer}
\pgfplotsset{
  legend style = {font=\ttfamily}
}
\pgfplotsset{
cycle list/Set1-5,
cycle multiindex* list={
mark list*\nextlist
Set1-5\nextlist
},
}

\usepackage{url}

\title{ANN-Benchmarks: A Benchmarking Tool for Approximate Nearest Neighbor Algorithms%
\footnote{The research of the first and third authors has received funding from the European Research Council under the European Union's 7th Framework Programme (FP7/2007-2013) / ERC grant agreement no. 614331.
A conference version of this work was published at SISAP'17 and is available at \url{http://dx.doi.org/10.1007/978-3-319-68474-1_3}.}
}

\author{Martin Aum\"{u}ller}
\author[2]{Erik Bernhardsson}
\author{Alexander Faithfull}
\affil{IT University of Copenhagen, Denmark, \{maau,alef\}@itu.dk}
\affil[2]{Better Inc., mail@erikbern.com}
\begin{document}

\maketitle
\begin{abstract}
    This paper describes \textsf{ANN-Benchmarks}, a tool for evaluating
    the performance of in-memory approximate nearest neighbor algorithms. It
    provides
    a standard interface for measuring the performance and quality achieved by nearest neighbor
    algorithms on different standard data sets. It
    supports several different ways of integrating $k$-NN algorithms, and
    its configuration system automatically tests a range of parameter settings
    for each algorithm.
    Algorithms are compared with respect to many different (approximate)
    quality
    measures, and adding more is easy and fast;
    the included plotting front-ends can visualise these as images, \LaTeX\
    plots, and websites with interactive plots.
    \textsf{ANN-Benchmarks} aims to provide a constantly updated overview of
    the current state of the art of $k$-NN algorithms.
    In the short term, this overview allows users to 
  	choose the correct $k$-NN algorithm and parameters for their similarity
  	search task;
	  in the longer term, algorithm designers will be able to use this overview
	  to test and refine automatic parameter tuning.
	The paper gives an overview of
    the system, evaluates the results of the benchmark,
    and points out directions for future work. Interestingly, very different approaches
	to $k$-NN search yield comparable quality-performance trade-offs. 
    The system is available at \url{http://ann-benchmarks.com}.
\end{abstract}

\section{Introduction}\label{sec:introduction}
Nearest neighbor search is one of the most fundamental tools in many areas of computer science, such as image recognition, machine learning, and computational linguistics.   For example, 
one can use nearest neighbor search on image descriptors such as MNIST \cite{Lecun98} to recognize handwritten digits, or one can find semantically
similar phrases to a given phrase by applying the \textsf{word2vec} embedding \cite{Mikolov13} and finding nearest neighbors. 
The latter can, for example, be used to tag articles on a news website and recommend new articles to readers that have shown an interest in a certain topic. In some cases, a generic  
nearest neighbor search under a suitable distance or measure of similarity offers surprising quality improvements \cite{BoytsovNMN16}.

In many applications, the data points are described by high-dimensional vectors, usually ranging from 100 to 1000 dimensions.
A phenomenon called the \emph{curse of dimensionality},
a consequence of several popular algorithmic hardness conjectures (see \cite{AlmanW15,Williams05}),
tells us that, to obtain the true nearest neighbors, we have to use either linear time (in the 
size of the dataset) or time/space that is exponential in the dimensionality of the dataset. In the case of \emph{massive} high-dimensional datasets, this rules out \emph{efficient} and \emph{exact} nearest neighbor search algorithms.

To obtain efficient algorithms, research has focused on allowing the returned neighbors to be an \emph{approximation} of the true nearest neighbors. Usually,
this means that the answer to finding the nearest neighbors to a query point is judged by how 
\emph{close} (in some technical sense) the result set is to
the set of true nearest neighbors.

There exist many different algorithmic techniques for finding approximate nearest neighbors.
Classical algorithms such as $kd$-trees \cite{kdtree} or M-trees \cite{CiacciaPZ97} can simulate this by terminating the search early, for example shown by 
Zezula et al. \cite{ZezulaSAR98} for M-trees. Other techniques \cite{hnsw,swgraph} build a graph from the dataset, where each vertex is associated with a data point, and a vertex is adjacent to its true nearest neighbors in the data set. Others involve projecting data points into a lower-dimensional space using hashing. 
A lot of research has been conducted with respect to locality-sensitive hashing (LSH) \cite{IndykM98}, but there exist many other 
techniques that rely on hashing for finding nearest neighbors; see \cite{wang} for a survey on the topic. 
We note that, in the realm of LSH-based techniques, algorithms guarantee sublinear query time, but solve a problem that is only distantly related to finding the $k$ nearest neighbors of a query point.
In practice, this could mean that the algorithm runs \emph{slower} than a linear scan through the data, and counter-measures have to be taken to avoid this behavior \cite{AhleAP17,Pham17}. 

Given the difficulty of the problem of finding nearest neighbors in high-dimensional spaces and the wide range of different solutions at hand, it is natural to ask how these algorithms perform in empirical settings. Fortunately, many of these 
techniques already have good implementations: see, e.g., \cite{flann,annoy,rpforest} for
tree-based, \cite{nmslib,kgraph} for graph-based, and \cite{falconn}
for LSH-based solutions.
This means that a new (variant of an existing) algorithm can show its worth by
comparing itself to the many previous algorithms on a collection of standard benchmark datasets with respect to a collection of quality measures.
What often happens, however, is that the evaluation of a new implementation is based
on a small set of competing algorithms and a small number of selected datasets. This approach poses problems for everyone involved: 
\begin{itemize}
	\item[(i)] \emph{The implementation's authors}, because competing implementations might be unavailable, they might use other conventions for input data and output of results, or the original paper might omit certain required parameter settings (and, even if these are available, exhaustive experimentation can take lots of CPU time).
	\item[(ii)] \emph{Their reviewers and readers}, because experimental results are difficult to reproduce and the selection of datasets and quality measures might appear selective.
\end{itemize}
This paper proposes a way of standardizing benchmarking for nearest neighbor
search algorithms, taking into account their properties and quality measures.
Our benchmarking framework provides a unified approach to experimentation and
comparison with existing work. The framework has already been used for
experimental comparison in other papers \cite{hnsw} (to refer to parameter
choice of algorithms) and algorithms have been contributed by the community,
e.g., by the authors of 
NMSLib \cite{nmslib} and FALCONN \cite{falconn}. An earlier version of our
framework is already widely used 
as a benchmark referred to from other websites, see, e.g., \cite{nmslib,falconn,annoy,rpforest,kgraph}.

\medskip

\noindent{\textbf{Related work.}} 
Generating reproducible experimental results is one of the greatest challenges in many areas of computer science, in particular in the machine learning community. As an example, \url{openml.org} \cite{openml2013} and \url{codalab.org} provide researchers with excellent platforms to share reproducible research results. 

The automatic benchmarking system developed in connection with the
\textsf{mlpack} machine learning library
\cite{mlpack2013,edel2014automatic} 
shares many characteristics with our framework: it automates the process of
running algorithms with preset parameters on certain datasets, and can
visualize these results. However, the underlying approach is very
different: it invokes whatever tools the implementations provide and parses
their standard output to extract result metrics. Consequently, the system
relies solely on the correctness of the algorithms' own implementations of
quality measures, and adding a new quality
measure would require a change in \emph{every single} algorithm implementation. Very recently,
Li et al. \cite{LiZSWZL16} presented a comparison of many approximate nearest neighbor algorithms, 
including many algorithms that are considered in our framework as well.  Their
approach is to take existing algorithm implementations and to heavily modify
them to fit a common style of query processing, in the process changing
compiler flags (and sometimes even core parts of the implementation). There
is no general framework, and including new features again requires manual
changes in each single algorithm.

Our benchmarking framework does not aim to replace these tools; instead, it
complements them by taking a different approach. We
only require that algorithms expose a simple programmatic interface for
building
data structures from training data and running queries. All the timing and quality measure
computation is conducted within our framework, which lets us add new metrics
without rerunning the algorithms, if the metric can be computed from the set
of returned elements. Moreover, we benchmark each implementation as intended 
by the author. That means that we benchmark \emph{implementations}, rather than \emph{algorithmic
ideas} \cite{KriegelSZ17}.

\medskip

\noindent{\textbf{Contributions.}} 
We describe our system for benchmarking approximate nearest neighbor algorithms with the general approach 
described in Section~\ref{sec:system:design}. The system allows for easy experimentation with $k$-NN algorithms, 
and visualizes algorithm runs in an approachable way. Moreover, in Section~\ref{sec:evaluation} we use our benchmark suite to overview the performance and quality of current state-of-the-art $k$-NN algorithms. This allows us to identify areas that already have competitive algorithms, to compare different methodological approaches to nearest neighbor search, but also to point out challenging datasets and metrics, where good implementations are missing or do not take full advantage of properties of the underlying metric. 
Having this overview has immediate practical benefits, as users can select the right combination of algorithm and parameters for their application. In the longer term, we expect that more algorithms will become able to tune their own parameters according to the user's needs, and our benchmark suite will also serve as a testbed for this automatic tuning.

\section{Problem Definition and Quality Measures}\label{sec:problem:desc}
We assume that we want to find nearest neighbors in a space $X$ with a distance measure $\text{dist}\colon X\times X \rightarrow \mathbb{R}$, for example the $d$-dimensional 
Euclidean space $\mathbb{R}^d$ under Euclidean distance ($l_2$ norm), or 
Hamming space $\{0,1\}^d$ under Hamming distance.

An algorithm $\mathcal{A}$ for nearest neighbor search builds a data structure DS$_\mathcal{A}$ for a data set $S \subset X$ of $n$ points. In a preprocessing phase, it creates DS$_\mathcal{A}$ 
to support the following type of queries: For a query point $q \in X$ and an integer $k$, return a \emph{result tuple} $\pi = (p_1, \ldots, p_{k'})$ of $k' \leq k$ distinct points from $S$ 
that are ``close'' to the query $q$. Nearest neighbor search algorithms generate $\pi$ by refining a set $C \subseteq S$ of 
candidate points w.r.t. $q$ by choosing the $k$ closest points among those using distance computations. The size of $C$ (and thus the number of distance computations) is denoted by $N$. We let $\pi^\ast = (p^\ast_1, \ldots, p^\ast_k)$ denote the tuple containing the true $k$ nearest neighbors for $q$ in $S$ (where ties are broken arbitrarily). We 
assume in the following that all tuples are sorted according to their distance to $q$.  

\subsection{Quality Measures}

We use different notions of ``recall'' as a measure of the quality of the result returned by the algorithm. 
Intuitively, recall is the ratio of the number of points in the result tuple that are true nearest neighbors to the number $k$ of true nearest neighbors. However, this intuitive 
definition is fragile when distances are not distinct or when we try to add a notion of approximation to it. To avoid these issues, we use 
the following distance-based definitions of recall and $(1+\varepsilon)$-approximative recall, that take the distance 
of the $k$-th  true nearest neighbor as threshold distance. 
\begin{align*}
\text{recall}(\pi, \pi^\ast) &= \frac{|\{p \text{ contained in $\pi$} \mid \text{dist}(p,q) \leq \text{dist}(p^\ast_k,q)\}|}{k}\\
\text{recall}_\varepsilon(\pi, \pi^\ast) &= \frac{|\{p \text{ contained in $\pi$} \mid \text{dist}(p,q) \leq (1 + \varepsilon) \text{dist}(p^\ast_k,q)\}|}{k}, \quad\text{for $\varepsilon > 0$.}
\end{align*}
(If all distances are distinct, $\text{recall}(\pi, \pi^\ast)$ matches the intuitive notion of recall.)

We note that (approximate) recall in high dimensions is sometimes criticised; see, for example, \cite[Section 2.1]{nmslib}. We investigate the impact of approximation as part of the evaluation in Section~\ref{sec:evaluation}, and plan to include other quality measures such as position-related measures \cite{ZezulaSAR98} in future work.

\begin{table}[t]
    \begin{tabular}{l l}
        \textbf{Name of Measure} & \textbf{Computation of Measure} \\ \hline \hline
        Index size of DS & Size of DS after preprocessing finished (in kB) \\
        Index build time DS & Time it took to build DS  (in seconds) \\ \hline
        Number of distance computations &  $N$ \\
        Time of a query & Time it took to run the query and generate result tuple $\pi$ \\[1em]
\end{tabular}
    \caption{Performance measures used in the framework.}
    \label{tab:performance:measures}
\end{table}
\subsection{Performance Measures}

With regard to the performance, we use the performance measures defined
in Table~\ref{tab:performance:measures}, which are divided 
into measures of the performance of the preprocessing step,
i.e., generation of the data structure, and measures of the performance 
of the query algorithm. With respect to the query performance, different
communities are interested in different cost values. Some rely on 
actual timings of query times, where others rely on the number of distance
computations. The framework can take both of
these measures into account. However, none of the currently included 
algorithms report the number of distance computations.

\section{System Design}\label{sec:system:design}
\textsf{ANN-Benchmarks} is implemented as a Python framework with several
different front-ends: one script for running experiments and a handful of
others for
working with and plotting results. It automatically downloads datasets when
they are needed and uses Docker build files to install algorithm implementations
and their dependencies.


This section gives only a high-level overview of the system; see
\url{http://ann-benchmarks.com} for more detailed technical
information.

\subsection{Algorithm implementations}

Each implementation is installed via a Docker build file. These files specify
how an implementation should be installed on a standard Ubuntu system by
building and installing its dependencies and code. ANN-Benchmarks requires that
this installation process also build Python wrappers for the implementation
to give the framework access to it. 

%

Adding support for a new algorithm implementation to \textsf{ANN-Benchmarks} is
as easy as
writing a Docker file to install it and its dependencies, making it available to
Python by writing a wrapper (or by reusing an existing one), and adding the
parameters to be tested to the configuration files. Most of the installation
scripts fetch the latest version of their library from its Git repository,
but there is no requirement to do this; indeed, installing several different
versions of a library would make it possible to use the framework for
regression testing.

We emphasise at this point that we are explicitly comparing algorithm
\emph{implementations}. Implementations make many different decisions that will
affect their performance and two implementations of the same algorithm
can have somewhat different performance characteristics \cite{KriegelSZ17}.
When implementations expose other quality measures -- such as the number of
distance computations, which are more suited for comparing algorithms on a
more abstract level -- our framework will also collect this information.

\medskip
\noindent{\textbf{Local mode.}}
Using Docker is ideal for evaluating the performance of well-tuned
implementations, but ANN-Benchmarks can also be used to help in the \emph{development} process.
To support this use case, the framework provides a \emph{local mode}, which
runs processes locally on the host system and not inside a Docker container.
This makes it much easier to build a pipeline solution to, for example,
automatically check how changes in the implementation influence its performance
-- in the standard Docker setup, each change would require the Docker container to be rebuilt.

\medskip
\noindent{\textbf{Algorithm wrappers.}}
To be usable by our system, each of the implementations to be tested
must have some kind of Python interface. Many libraries already
provide their own Python wrappers, either written by hand or automatically
generated using a tool like SWIG; others are implemented partly or entirely
in Python.



To bring implementations that do not provide a Python interface into the
framework,
we specify a simple text-based protocol that supports the few operations we
care about: parameter configuration,
sending training data, and running queries. The framework comes with a wrapper
that communicates with external programs using this protocol.
In this way, experiments can be run in
external front-end processes implemented in any programming language.


The protocol has been designed to be easy to implement. Every message is a line
of text that will be split into tokens according to the rules of the POSIX
shell, good implementations of which are available for most programming
languages.
The protocol is flexible and extensible: front-ends are free to
include extra information in replies, and they can also implement special
configuration options that cause them to diverge from the protocol's basic
behaviour. As an example, we provide a simple C implementation that
supports an alternative query mode in which parsing and preparing a
query data point and running a query are two different commands. (As the
overhead of parsing a string representation of a data point is introduced by
the use of the protocol, removing it makes the timings more representative%
.)

The use of a plaintext protocol necessarily adds some overhead, but this is
often not terribly significant -- indeed, we have found that a na\"ive linear
search implemented in Java and invoked using the protocol is \emph{faster} than
a na\"ive Python implementation.

\subsection{Datasets and ground truth}

By default, the framework fetches datasets on demand from a remote server.
These dataset files contain, in HDF5 format,
the set of data points,
the set of query points,
the distance metric that should be used to compare them,
a list of the true nearest $k = 100$ neighbours for each query point,
and a list of the distances of each of these neighbours from the query point.

The framework also includes a script for generating dataset files from the
original datasets. Although using the precomputed hosted versions is normally
simpler, the script can be used to, for example, build a dataset file with a
different value of $k$, or to convert a private dataset for the framework's
use.

Most of the datasets use as their query set a pseudorandomly-selected set of
ten thousand entries separated from the rest of the training data; others have
separate query sets. The dataset file generation script makes this decision.

\subsection{Creating algorithm instances}

\begin{figure}[t!]
\begin{lstlisting}[language=yaml]
float:
  euclidean:
    megasrch:
      docker-tag: ann-benchmarks-megasrch
      module: ann_benchmarks.algorithms.MEGASRCH
      constructor: MEGASRCH
      base-args: ["@metric"]
      run-groups:
        shallow-point-lake:
          args: ["lake", [100, 200]]
          query-args: [100, [100, 200, 400]]
        deep-point-ocean:
          args: ["sea", 1000]
          query-args: [[1000, 2000], [1000, 2000, 4000]]
\end{lstlisting}
\caption{An example of a fragment of an algorithm configuration file.}
\label{fig:overview:example}
\end{figure}

After loading the dataset, the framework moves on to creating the algorithm
instances. It does so based on a YAML configuration file that specifies a
hierarchy of dictionaries: the first level specifies the point type, the second
the distance metric, and the third each algorithm implementation to be tested.
Each implementation
gives the name of its wrapper's Python constructor; a number of other entries
are then expanded to give the arguments to that constructor.
Figure~\ref{fig:overview:example} shows an example of this configuration file.

The \texttt{base-args} list consists of those arguments that should be
prepended to every invocation of the constructor. Figure~\ref{fig:overview:example}
also shows one of the special keywords, \texttt{"@metric"}, that is used to
pass one of the framework's configuration parameters to the constructor.

Algorithms must specify one or more ``run groups``, each of which will be
expanded into one or more lists of constructor arguments. The \texttt{args}
entry completes the argument list, but not directly:
instead, the Cartesian product of all of its entries is used to generate
\emph{many} lists of arguments. 
Another entry, \texttt{query-args}, is expanded in the same way as \texttt{args},
but each argument list generated from it is used to reconfigure the query
parameters of an algorithm instance after its internal data structures have
been built. This allows built data structures to be reused, greatly reducing
duplicated work.

As an example,
the \texttt{megasrch} entry in 
Figure~\ref{fig:overview:example} expands into
three different algorithm instances: \texttt{MEGASRCH("euclidean", "lake", 100)},
\texttt{MEGASRCH("euclidean", "lake", 200)}, and \texttt{MEGASRCH("euclidean", "sea", 1000)}. Each of
these will be trained once and then used to run a number of experiments: the
first two will run experiments with each of the query parameter groups \texttt{[100, 100]},
\texttt{[100, 200]}, and \texttt{[100, 400]} in turn, while the last will run
its experiments with the query parameter groups
\texttt{[1000, 1000]}, \texttt{[1000, 2000]}, \texttt{[1000, 4000]},
\texttt{[2000, 1000]}, \texttt{[2000, 2000]}, and \texttt{[2000, 4000]}.


\subsection{The experiment loop}

\begin{figure}[t!]
\hspace{-5em}
\includegraphics[width=1.3\textwidth]{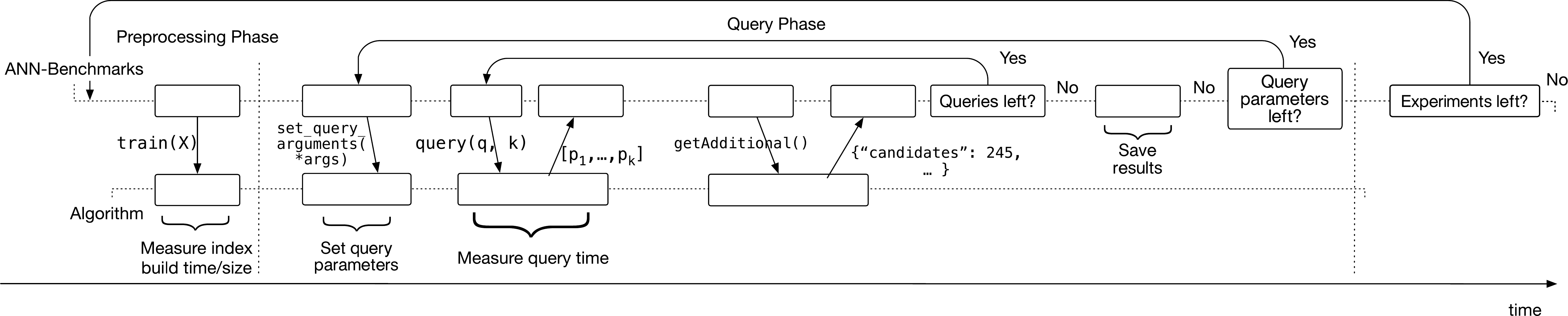}
\caption{Overview of the interaction between ANN-Benchmarks and an algorithm
instance under test.}
\label{fig:overview:experiment}
\end{figure}

Once the framework knows what instances should be run, it moves on to the
experiment
loop, shown in Figure~\ref{fig:overview:experiment}.
The loop consists of two phases.
In the \emph{preprocessing phase}, an algorithm instance builds an index data
structure for the dataset $X$. The loop then transitions to the \emph{query
phase}, in which query points are sent one by one to the algorithm instance.
For each query point, the instance returns (at most) $k$ data points; after
answering a query, it can also report any extra
information it might have, such as the number of candidates considered, i.e., 
the number of exact distances computed. The instance
is then reconfigured with a new set of query parameters, and the query set is
run repeatedly, until no more sets of these parameters remain.

Each algorithm instance is run in an isolated Docker container. This makes it
easy to clean up after each run: simply terminating the container 
takes care of everything. Moving experiments out of the main process
also gives us a simple and
implementation-agnostic way of computing the memory usage of an implementation:
the subprocess records its total memory consumption before and after
initialising the algorithm instance's data structures and compares the two
values.

The complete results of each run are written to the host by mounting
part of the file system into the Docker container. The main process performs a
blocking, timed wait on the container, and will terminate it if
the user-configurable timeout is exceeded before any results are available.

\medskip

\noindent\textbf{Dataset size.} In its current form, ANN-Benchmarks supports
benchmarking \emph{in-memory} nearest-neighbor algorithms. In particular, the dataset is kept 
in memory by ANN-Benchmarks when running experiments. This has to be taken into account when choosing datasets to include into the framework. In practice, this means that the framework can handle datasets with millions of points of dimensionality up to a few thousand dimensions.

\subsection{Multi-threading and batched queries}

The experiment loop is, by default, run on a single CPU in a single thread. The single-threaded mode is enforced when the Docker container is started, using the Linux kernel's \texttt{cpusets} capabilities to restrict access to the system's resources.  
Running on a single CPU makes the comparison between implementations fairer, since all implementations run on the same grounds.   

However, parallelizing single queries or using parallelism over queries is an important topic. In fact, in many real-world systems that deploy nearest-neighbor algorithms, queries can be batched together. This means that the data structure receives a sequence of queries all at once, and returns results for all of the queries contained in that sequence.
This enables many interesting approaches to parallelization that would not be possible when running single queries.

ANN-Benchmarks supports these systems with a \emph{batch mode}, in which the whole set of queries is given to the implementation wrapper at once. In this mode, all resources of
the host system are made available to the Docker container.
The behaviour of the experiment loop diverges slightly from Figure~\ref{fig:overview:experiment} in batch mode. Batch queries do not return a sequence of tuples containing answers to the individual queries; instead, these results are obtained via an additional method, akin to Figure~\ref{fig:overview:experiment}'s \texttt{getAdditional()} method. This allows an algorithm to return the result of a batch query as an opaque internal data structure; this will stop the clock, and the additional call can then transform that data structure into Python objects without that transformation imposing a performance penalty.

Batch query mode is particularly useful for running nearest-neighbor algorithms on a GPU. In this context, transferring a single query point to the GPU memory and getting the result of the query from the GPU can be a dominating part of the query time, as we will see in Section~\ref{sec:evaluation}.

\subsection{Results and metrics}


For each run, we store the full name -- including the parameters -- of the
algorithm instance, the time it took to build its index data structure, and the
results of every query: the near neighbours returned by the algorithm, the time
it took to find these, and their distances from the query point, along with any
additional information the implementation might have exposed.
(To avoid affecting the timing of algorithms that do not indicate
the distance of a result, the experiment loop independently re-computes
distance values after the query has otherwise finished.)

The results of each run are stored in a separate HDF5 file in a directory
hierarchy that encodes
part of the framework's configuration. 
Keeping runs in separate files makes them easy to enumerate and easy to re-run,
and individual
results -- or sets of results -- can easily be shared to make results more
transparent.

Metric functions are passed the ground truth and the results for a particular
run; they can then compute their result however they see fit. Adding a new
quality metric is a matter of writing a short Python function and adding it to
an internal data structure; the plotting scripts query this data structure and
will automatically support the new metric.

\subsection{Frontend}\label{sec:frontend}

\begin{figure}[t]
\centering
\includegraphics[width=0.9\textwidth]{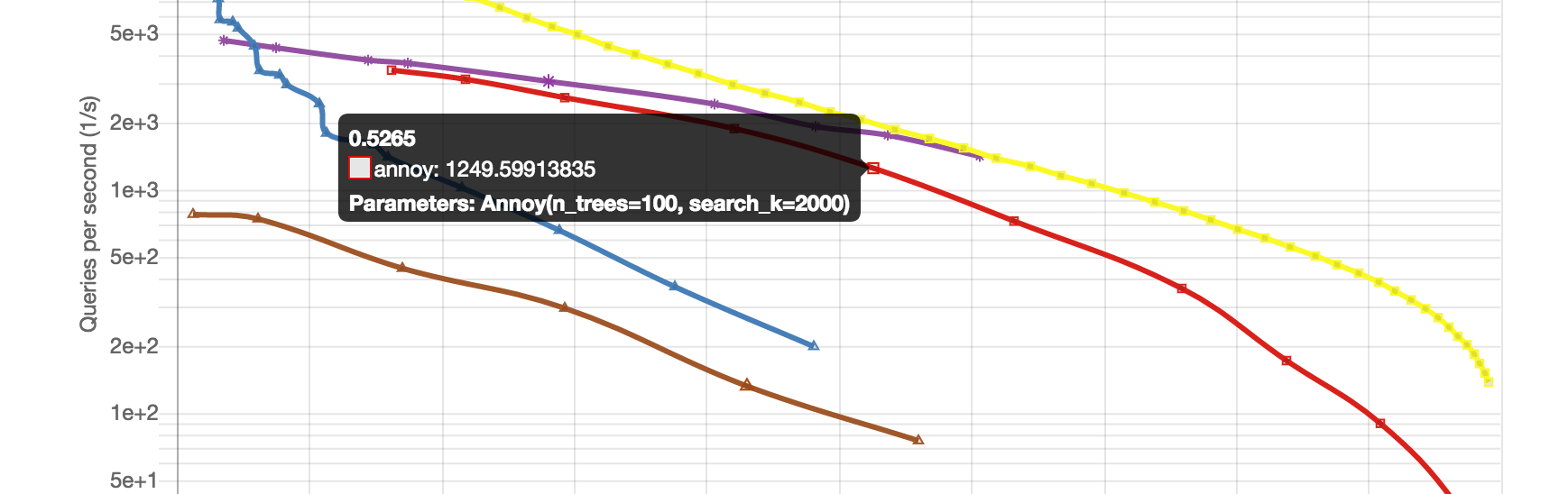}
\caption{Interactive plot screen from framework's website (cropped). 
Plot shows ``Queries per second'' ($y$-axis, log-scaled) against ``Recall'' ($x$-axis, not shown). 
Highlighted data point corresponds to a run of \textsf{Annoy} with parameters 
as depicted, giving about 1249 queries per second for a recall of about $0.52$.}
\label{fig:interactive:plot}
\end{figure}

\textsf{ANN-Benchmarks} provides two options to evaluate the results of the experiments: a script to generate individual plots using Python's \textsf{matplotlib} and a script to generate a website that summarizes the results and provides interactive plots with the option to export the plot as {\LaTeX} code using \textsf{pgfplots}. See Figure~\ref{fig:interactive:plot} for an example. Plots depict the Pareto frontier over all runs of an algorithm; this gives an
immediate impression of the algorithm's general characteristics, at the cost of
concealing some of the detail. When more detail is desired, the scripts can
also produce scatter plots.

As batch mode goes to greater lengths to reduce overhead than the normal query
mode and exposes more of the system's resources to the implementation being
tested, results obtained in batch mode are always presented separately by the
evaluation scripts to make the comparisons fairer.

\section{Evaluation}\label{sec:evaluation}
In this section we present a short evaluation of our findings from running benchmarks in the benchmarking framework. After discussing the evaluated implementations and datasets, we will present four questions that we intended to answer using the framework. Subsequently, we will discuss the answers to these questions and present some observations regarding the build time of indexes and their ability to answer batched queries. At the end of this section, we present a summary of our findings. 

\begin{table}[t]
    \begin{tabular}{l l}
        \textbf{Principle} & \textbf{Algorithms} \\ \hline \hline 
        $k$-NN graph & \texttt{KGraph (KG)} \cite{kgraph}, \texttt{SWGraph (SWG)} \cite{swgraph,nmslib}, \texttt{HNSW} \cite{hnsw,nmslib}, \texttt{PyNNDescent (NND)} \cite{pynndescent}, PANNG \cite{ngt,Iwasaki16}  \\
        tree-based & \textit{FLANN} \cite{flann}, \textit{BallTree (BT)} \cite{nmslib}, \texttt{Annoy (A)} \cite{annoy}, \texttt{RPForest (RPF)} \cite{rpforest}, \texttt{MRPT} \cite{Hyvonen2016,mrpt} \\
        LSH & \texttt{FALCONN (FAL)} \cite{falconn}, \textit{MPLSH} \cite{mplsh,nmslib}
        \\
        
    other 
              & \texttt{Multi-Index Hashing (MIH)} \cite{mihalgo} (exact Hamming search), \\
              & \texttt{FAISS-IVF (FAI)} \cite{faiss} (inverted file)
    \end{tabular}
    \caption{Overview of tested algorithms (abbr. in parentheses). Implementations in \textit{italics} have ``recall'' as quality measure
     provided as an input parameter.} 
    \label{tab:algorithms}
    
\end{table}
\begin{table}[t]
\begin{tabular}{l l l l}
    \textbf{Dataset} & \textbf{Data/Query Points} & \textbf{Dimensionality} & \textbf{Metric} \\ \hline \hline
    \textsf{SIFT} & 1\,000\,000 / 10\,000 & 128 & Euclidean \\
    \textsf{GIST} & 1\,000\,000 / 10\,000 & 960 & Euclidean \\
    \textsf{GLOVE} & 1\,183\,514 / 10\,000 & 100 & Angular/Cosine \\ 
    \textsf{NYTimes} & 234\,791 / 10\,000 & 256 & Euclidean \\
    \textsf{Rand-Euclidean} & 1\,000\,000 / 10\,000 & 128 & Angular/Cosine \\
    \textsf{SIFT-Hamming} & 1\,000\,000 / 1\,000& 256 & Hamming \\
    \textsf{Word2Bits} & 399\,000 / 1\,000 & 800 & Hamming \\
\end{tabular}
\caption{Datasets under consideration.}
\label{tab:datasets}

\end{table}

\medskip

\noindent{\textbf{Experimental setup.}} All experiments were run in Docker
containers on \emph{Amazon EC2} 
\emph{c5.4xlarge} instances that are equipped with Intel Xeon Platinum 8124M CPU (16 cores available, 3.00 GHz, 25.0MB Cache)  and 
32GB of RAM running Amazon Linux. We ran a single experiment multiple times to verify that performance was reliable, and compared the experiments results with a 4-core Intel Core i7-4790 clocked at 3.6 GHz with 32GB RAM. While the latter was a little faster, the relative order of algorithms remained stable. For each parameter setting and dataset, the algorithm was given 
five hours to build the index and answer the queries.  

\medskip

\noindent{\textbf{Tested Algorithms.}} Table~\ref{tab:algorithms} summarizes the algorithms that are used in the evaluation; see the references provided for details. The framework has support for more implementations and many of these were included 
in the experiments, but they
turned out to be either non-competitive or too similar to other implementations.\footnote{For example, the framework contains three different 
implementations of HNSW: the original one from NMSlib, a standalone variant inspired by that one, and an implementation in FAISS that is 
again inspired by the implementation in NMSlib. The first two implementations perform almost indistinguishably, while the implementation provided in FAISS was a bit slower. For the sake of brevity, we also omit the two random projection forest-based methods \texttt{RPForest} and \texttt{MRPT} since they were always slower than \texttt{Annoy}. } The scripts that set up the framework automatically fetch the most current version found in each algorithm's repository.

In general, the implementations under evaluation can be separated into three main algorithmic principles: graph-based, tree-based, and hashing-based algorithms. \emph{Graph-based algorithms} build a graph in which vertices are the points in the dataset and edges connect vertices that are true nearest neighbors of each other, forming the so-called $k$-NN graph. Given the query point, close neighbors are found by traversing the graph in a greedy fashion, subject to the actual implementation \cite{kgraph,swgraph,hnsw,pynndescent,Iwasaki16}.  \emph{Tree-based
algorithms} use a collection of trees as their data structure. In these trees, each node splits the dataset into subsets that are then processed in the children of the node. If the dataset associated with a node is small enough, it is directly stored in the node which is then a leaf in the tree.  For example, \texttt{Annoy} \cite{annoy} and \texttt{RPForest} \cite{rpforest} choose in each node a random hyperplane to split the dataset. Given the query point, the collection of trees are traversed to obtain a set of
candidate points from which the closest to the query are returned. \emph{Hashing-based algorithms} apply hash functions such as locality-sensitive hashing \cite{IndykM98} to map data points to hash values. At query time, the query point is hashed and keys colliding with it, or not too far from it using the
multi-probe approach \cite{mplsh}, are retrieved. Among them, those closest to the query point are returned. Different implementations are mainly characterized by the underlying locality-sensitive hash function that is being used.

\medskip

\noindent{\textbf{Datasets.}} The datasets used in this evaluation are summarized in Table~\ref{tab:datasets}. More informations on these datasets and results for other datasets are found on the framework's website. The \textsf{NYTimes} dataset was generated by building tf-idf descriptors from the bag-of-words version, and embedding them into a lower dimensional space using the Johnson-Lindenstrauss Transform \cite{JohnsonL86}. The Hamming space version of \textsf{SIFT}
was generated by applying Spherical Hashing \cite{HeoLHCY15}  using the implementation provided by the authors of \cite{HeoLHCY15}. The dataset \textsf{Word2Bits} comes from 
the quantized word vector approach described in \cite{Lam18} using the top-400\,000 words in the English Wikipedia from 2017.

The dataset \textsf{Rand-Euclidean} is generated as follows: Assume that we want to generate a dataset with $n$ data points, $n'$ query points, and are interested in finding the $k$ nearest neighbors for each query point. For an even dimension $d$, we generate $n - k\cdot n'$ data points of the form $(v, \mathbf{0})$, where $v$ is a random unit length vector of dimension $d/2$, and $\mathbf{0}$ is the vector containing $d/2$ $0$ entries. We call the first $d/2$ components the \emph{first part} and the following $d/2$ components the \emph{second part} of the vector.
From these points, we randomly pick $n'$ points $(v_1,\ldots,v_{n'})$. For each point $v_i$, we replace its second part with a random vector of length $1/\sqrt{2}$. The resulting point is the query point $q_i$. For each $q_i$, we insert $k$ random points at varying distance increasing from $0.1$ to $0.5$ to $q_i$ into the original dataset. The idea behind such a dataset is that the vast majority of the dataset looks like a random dataset with little structure for the algorithm to exploit, while
each query point has $k$ neighbors that are with high probability well separated from the rest of the data points. This means that the queries are easy to answer locally, but they should be difficult to answer if the algorithm wants to exploit a global structure.
 
\medskip
 
\noindent{\textbf{Parameters of Algorithms.}} Most algorithms do not allow the user to explicitly specify a quality target---in fact, only 
three implementations from Table~\ref{tab:algorithms} provide ``recall'' as an input parameter. We used our framework to test many parameter settings at once. The detailed settings tested for each algorithm can be found on the framework's website. 

\medskip 

\noindent{\textbf{Status of FALCONN.}} While preparing this full version, we noticed that \texttt{FALCONN}'s performance has drastically decreased in the latest versions. We communicated this to the authors of \cite{falconn}, who are now working on a fix; however, they asked us to disregard \texttt{FALCONN} for this submission. We plan to include it in a revised version.

\subsection{Objectives of the Experiments}
We used the benchmarking framework to find answers to the following questions:

\noindent\textbf{(Q1) Performance.} Given a dataset, a quality measure and a number $k$ of nearest neighbors to return, how do algorithms compare to each other with respect 
        to different performance measures, such as query time or index size?\\
\noindent\textbf{(Q2) Robustness.} Given an algorithm $\mathcal{A}$, how is its performance 
        and result quality influenced by the dataset and the number of returned neighbors?\\
\noindent\textbf{(Q3) Approximation.} Given a dataset, a number $k$ of nearest neighbors to return, 
        and an algorithm $\mathcal{A}$, how does its performance improve when the returned neighbors can be an approximation? Is the effect comparable for different algorithms? \\
\noindent\textbf{(Q4) Embeddings.} Equipped with a framework with many different datasets and distance metrics, we can try interesting combinations. How do algorithms targeting Euclidean space or Cosine similarity perform in, say, Hamming space? How does replacing the internals of an algorithm with Hamming space related techniques improve its performance?

The following discussion is based on a combination of the plots found on the
framework's website; see the website for more complete and up-to-date results.
\begin{figure}[t!]
\input{plot-performance}
\end{figure}

\subsection{Discussion}

\noindent{\textbf{(Q1) Performance.}} Figure~\ref{plot:performance} shows the relationship between an algorithm's achieved recall and the number of queries it can answer per second (its QPS) on the two datasets \textsf{GLOVE} (Cosine similarity) and \textsf{SIFT} (Euclidean distance) for $10$- and $100$-nearest neighbor queries.

For \textsf{GLOVE}, we observe that the graph-based algorithms clearly outperform the tree-based approaches. It is noteworthy that all implementations, except \texttt{FLANN}, 
achieve close to perfect recall. Over all recall values, \texttt{HNSW}
is fastest. However, at high recall values it is closely matched by \texttt{KGraph}. \texttt{FAISS-IVF} comes in at third place, only losing to the 
other graph-based approaches at very high recall values. For 100 nearest neighbors, the picture is very similar. We note, however, that 
the graph-based indexes were not able to build indexes for nearly perfect
recall values within 5 hours. 

On \textsf{SIFT}, all tested algorithms 
can achieve
close to perfect recall. Again,
the graph-based algorithms are fastest; they are followed by \texttt{Annoy} and
\texttt{FAISS-IVF}. \texttt{FLANN} and \texttt{BallTree} are at the end.
In particular, \texttt{FLANN} was not able to finish its auto-tuning for high
recall values within 5 hours. 

Very few of these algorithms can tune themselves to produce a particular recall
value. In particular, almost all of the fastest algorithms on the \textsf{GLOVE} dataset
expose many parameters, leaving the user to find the combination that works
best. The
\texttt{KGraph} algorithm, on the other hand, uses only a single parameter,
which---even in its ``smallest'' choice---still gives high recall on \textsf{GLOVE} and \textsf{SIFT}.
\texttt{FLANN} manages to tune itself for a particular recall value well. However, at 
high recall values, the tuning does not complete within the time
limit, especially with 100-NN.

\begin{figure}[t!]
\input{plot-index-size}
\end{figure}

Figure~\ref{plot:index:size} relates an algorithm's performance to its index
size. (Note that here down and to the right is better.) High recall can be achieved with small indexes by probing many points;
however, this probing is expensive, and so the QPS drops dramatically.
To reflect this performance cost, we scale the size of the index by the QPS it
achieves for a particular run.
This reveals that, on \textsf{SIFT}, most implementations perform similarly under 
this metric. \texttt{HNSW} is best (due to the QPS it achieves), but most of the other 
algorithm achieve similar cost. In particular, \texttt{FAISS-IVF} and \texttt{FLANN} do well. \texttt{NND}, \texttt{Annoy}, and \texttt{BallTree} achieve their QPS at the cost of relatively large indexes, reflected in a rather large gap between them and their competition. On \textsf{GLOVE}, we see a much wider spread of index size performance. Here, \textsf{FAISS-IVF} and \textsf{HNSW} perform nearly indistinguishably. Next follow the other graph-based algorithms, with \texttt{FLANN} among
them. Again, \texttt{Annoy} and \texttt{BallTree} perform worst in this measure.

\begin{figure}[t!]
    \begin{tikzpicture}[every mark/.append style={mark size=2pt}]
        \begin{groupplot}[group style = {group size = 1 by 1, group name = hamming, vertical sep=2.1cm}, grid = both, grid style={line width=.1pt, draw=gray!30},
    major grid style={line width=.2pt,draw=gray!50},height=4.5cm, width=\textwidth,
              ymin = 0, ymax = 10000, xtick = {0, 0.25, 0.5, 0.75, 1}, max space between ticks=20]

        \nextgroupplot[
            ylabel={QPS (1/s)},
            ylabel style ={overlay},
            ymode = log,
            yticklabel style={/pgf/number format/fixed,
                              /pgf/number format/precision=3},
            legend style = { at = { (.95, 1.25 ) }},
            legend columns = 7
            ]
 \addplot coordinates {
                                    (0.95604, 8536.03210325)
                                    (0.98129, 8335.10993042)
                                    (0.98944, 6528.86173483)
                                    (0.99451, 5649.79063345)
                            };
            \addlegendentry{A};
                                \addplot coordinates {
                                                (0.85208, 2543.2123768)
                                    (0.99312, 1662.87808506)
                            };
            \addlegendentry{FAI-IVF};
                                \addplot coordinates {
                                    (0.26139, 3346.00025193)
                                    (0.47712, 3122.91843795)
                                    (0.56114, 2214.12838818)
                                    (0.61553, 1610.65353579)
                                    (0.65768, 1325.49508662)
                                    (0.69465, 1127.4738886)
                                    (0.72153, 971.657818132)
                                    (0.74107, 862.728281433)
                                    (0.75764, 773.762333342)
                                    (0.7753, 702.51147235)
                                    (0.79559, 601.31894065)
                                    (0.8124, 524.013305675)
                                    (0.82584, 463.936347326)
                                    (0.84262, 376.057132369)
                                    (0.86225, 254.30851466)
                                    (0.86912, 192.053011214)
                            };
            \addlegendentry{HNSW};
                                \addplot coordinates {
                                    (0.7305, 5971.90560175)
                                    (0.9448, 4498.41821382)
                                    (1.0, 3793.642952)
                            };
            \addlegendentry{PANNG};
                                \addplot coordinates {
                                                (0.98645, 2546.41358421)
                                    (0.99265, 2431.62408077)
                            };
            \addlegendentry{NND};
                                \addplot coordinates {
                                                (0.99558, 2504.43782999)
                            };
            \addlegendentry{KGraph};
                                \addplot coordinates {
                                                (0.63882, 3437.15510067)
                                    (0.66099, 3358.16465664)
                                    (0.66694, 3329.5108387)
                                    (0.66933, 2994.97410649)
                                    (0.67083, 2993.13284936)
                                    (0.68122, 2667.73096796)
                                    (0.7344, 2032.26670322)
                                    (0.7434, 1494.46493019)
                                    (0.90251, 1340.78127998)
                                    (0.92269, 1225.19979109)
                                    (0.92904, 702.45871537)
                                    (0.93346, 476.740636167)
                                    (0.94332, 285.781290233)
                                    (0.96354, 160.545298785)
                                    (0.98803, 88.375436211)
                                    (0.99791, 53.0236556907)
                            };

            \addlegendentry{ SWG};
    \end{groupplot}
\end{tikzpicture}
    \caption{Recall-QPS (1/s) tradeoff - up and to the right is better; \textsf{Rand-Euclidean} with 10-NN.}
    \label{plot:random}
\input{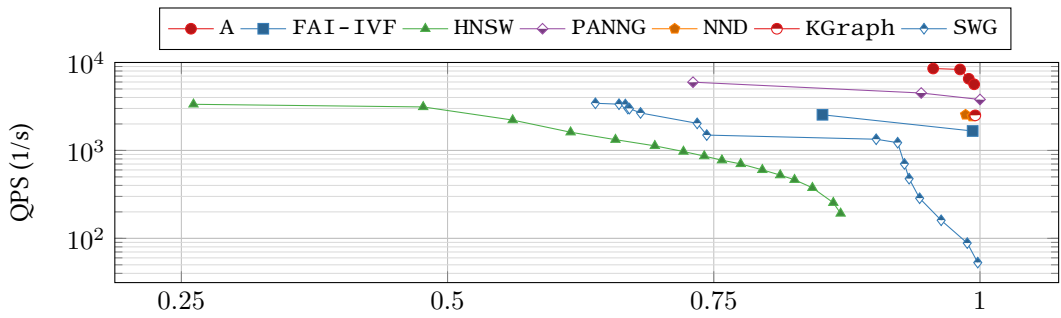}
\end{figure}

\noindent{\textbf{(Q2) Robustness.}} Figure~\ref{plot:random} plots recall against QPS on the dataset \textsf{Rand-Euclidean}. Recall from our earlier discussion of datasets that this dataset contains easy queries, but requires an algorithm to exploit the local structure instead of some global structure of the data structure, cf.~\textbf{Datasets}. We see very different behavior than before: there is a large difference between different graph-based approaches. While \texttt{PANNG}, \texttt{KGraph},
\texttt{NND}
can solve the task easily with high QPS, both \texttt{HNSW} and \texttt{SWG} fail in this task. This means that the ``small-world'' structure of these two methods \emph{hurts} performance on such a dataset. In particular, no tested parameter setting for \texttt{HNSW} achives recall beyond .86. \texttt{Annoy} performs best at exploiting the local structure of the dataset and is the fastest algorithm. The dataset is also easy for \texttt{FAISS-IVF}, which also has very good performance.    

Let us turn our focus to how the algorithms perform on a wide variety of datasets. Figure~\ref{plot:robustness} plots recall against QPS for \texttt{Annoy}, \texttt{FAISS-IVF}, and \texttt{HNSW} over a range of datasets. Interestingly, implementations agree on the ``difficulty'' of a dataset most of the time, i.e., the relative order
of performance is the same among the algorithms. Notable exceptions are \textsf{Rand-Euclidean}, which is very easy for \texttt{Annoy} and \texttt{FAISS-IVF}, but difficult for \texttt{HNSW} (see above), and \textsf{NYTimes}, where \texttt{FAISS-IVF} fails to achieve recall above .7 for the tested parameter settings. Although all algorithms take a performance hit for high recall values, \texttt{HNSW} is least affected. On the other hand, \texttt{HNSW} shows the biggest slowdown in answering 100-NN
compared to 10-NN queries among the different algorithms.

\begin{figure}[t!]
\input{plot-epsilon}
\end{figure}
\begin{figure}[t!]
    \begin{tikzpicture}[every mark/.append style={mark size=1.5pt}]
        \begin{groupplot}[group style = {group size = 2 by 1, group name = hamming}, height=6cm, width=.55\textwidth,
            ylabel style = {yshift=-1.5ex}, xlabel style = {yshift=1.5ex}, ymin = 10, ymax = 100000, xtick = {0, 0.25, 0.5, 0.75, 1}, max space between ticks=20, grid = both, grid style={line width=.1pt, draw=gray!30},
    major grid style={line width=.2pt,draw=gray!50},]
        \nextgroupplot[
            xlabel={Recall},
            ylabel={QPS (1/s)},
            ymode = log,
            yticklabel style={/pgf/number format/fixed,
                              /pgf/number format/precision=3},
            ylabel style = {overlay},
            legend style = {at = {(1.95, 1.33)}},
            legend columns = 7
            ]
    
  \addplot coordinates { (0.4635, 5284.00940823) (0.5692, 4253.26704099) (0.703, 3246.96596352) (0.7304, 2258.25719761) (0.8625, 2011.3806931) (0.8794, 1563.44211155) (0.9389, 1327.17573489) (0.9495, 1120.3554961) (0.9576, 959.986743347) (0.9773, 864.639864636) (0.9815, 778.103664688) (0.9856, 703.332201443) (0.9956, 445.6169935)  };
    
        \addlegendentry{A};
\addplot coordinates { (0.3527, 15997.6810002) (0.5273, 9259.991743) (0.5275, 8562.73987816) (0.7329, 4732.02266332) (0.8964, 2265.4182024) (0.9396, 1879.99306147) (0.9401, 1865.32892162) (0.9513, 1369.51051874) (0.98, 1096.61743924) (0.9916, 898.877020934) (0.9939, 661.351847741) (0.9986, 526.283667668) (0.9987, 310.467488401)
                            };            
        \addlegendentry{A (Ham.)};

 \addplot coordinates { (0.5656, 8443.32071825) (0.6159, 5272.34582272) (0.8876, 4822.99599723) (0.9077, 4397.71134274) (0.9382, 4061.65442803) (0.9678, 3105.42655058) (0.9906, 2135.96678848) (0.9954, 2082.24934146) (0.9975, 1701.32579407) (0.998, 1428.45310491) (0.9993, 1317.66698742) 
                            };
            
        \addlegendentry{FAI-IVF};

        \addplot coordinates { (0.4359, 7403.79907044) (0.9252, 3981.74265035) (0.9569, 3336.41230422) (0.9898, 1909.47452284) (0.9997, 853.188392678)  }; 

        \addlegendentry{PANNG}; 

        \addplot[mark=*,mark size=4, smooth]  coordinates { (1.0, 469.865268002) };
        \addlegendentry{MIH};
 
\addplot coordinates { (0.084543, 5584.3854113) (0.119391, 4446.88200639) (0.174152, 3247.82055985) (0.189256, 2266.7971737) (0.278427, 1939.72103986) (0.299902, 1481.26194653) (0.319946, 1255.66460079) (0.375522, 1240.63373154) (0.401737, 1022.7544419) (0.425452, 934.016037545) (0.483465, 765.689211275) (0.512056, 688.535908575) (0.536908, 649.434006022) (0.631789, 380.240612309) (0.659541, 375.267750327) (0.681717, 360.018396517) (0.762977, 231.717874622) (0.781039, 217.512564304) (0.849984, 130.316700287) (0.865492, 127.74095228) (0.943076, 58.7219372326) (0.976695, 32.1364902735) (0.992438, 17.6399500268) };
\addlegendentry{A (Eucl.)}

        \nextgroupplot[
            xlabel={Recall},
            ymode = log,
            yticklabel style={/pgf/number format/fixed,
                              /pgf/number format/precision=3},
            ]
        \addplot coordinates { (0.5381, 2344.24608664) (0.5453, 2256.25158018) (0.556, 2052.11424859) (0.566, 1980.58655868) (0.5889, 1568.75288697) (0.5966, 1527.57391571) (0.6963, 1318.82251583) (0.7125, 1234.13135102) (0.7367, 1027.41057167) (0.7968, 878.880997787) (0.8095, 840.990861654) (0.8276, 726.706827274) (0.8696, 563.523689053) (0.8817, 550.165805122) (0.8933, 502.060885971) (0.9394, 294.433655722) (0.9442, 282.678680381) (0.9634, 178.055121004) (0.9652, 177.793829212) (0.9803, 109.472143695) (0.9809, 105.277298478) (0.9924, 54.3614930991) };    

\addplot coordinates { (0.0474, 15717.8928907) (0.051, 15080.2098277) (0.0561, 13700.2495525) (0.0885, 11447.9611332) (0.09, 11108.4732386) (0.0948, 10404.2685665) (0.1506, 7436.14227993) (0.1592, 7167.71651121) (0.2991, 3563.24600587) (0.3055, 3480.92642034) (0.4557, 1850.66718909) (0.4627, 1839.80875096) (0.4706, 1803.02614757) (0.5582, 1078.65732584) (0.6418, 911.293704273) (0.6456, 902.540249323) (0.735, 475.921793358) (0.7799, 437.332701607) (0.8402, 362.472237538) (0.8697, 246.308199307) (0.8863, 231.811584555) (0.9174, 206.327224444) (0.947, 128.801289615) (0.9641, 120.94416928) (0.9716, 114.397820544) (0.99, 59.1350919818) };
            
\addplot coordinates { (0.5256, 10285.7282989) (0.5735, 6658.80918512) (0.6285, 5537.95614848) (0.6592, 4087.48971867) (0.8307, 3926.11133472) (0.8486, 3460.85531554) (0.8936, 3234.79068318) (0.9127, 2206.55057075) (0.9586, 1746.04504087) (0.9695, 1072.37906185) (0.9706, 1055.01820491) (0.9777, 945.656854704) (0.9826, 707.27557432) (0.9834, 622.895374379) (0.9869, 544.948779307) (0.9882, 337.565500184) (0.9926, 331.978183791) };
            
\addplot coordinates { (0.0067, 4806.44847079) (0.0082, 4634.78978652) (0.0425, 4528.31231646) (0.1276, 4198.30918696) (0.7862, 2065.51654854) (0.9387, 1414.67615067) (0.9912, 517.717709266) };

        \addplot[mark=*,mark size=4, smooth]  coordinates {(1.0, 19.9397) };
\end{groupplot}
\end{tikzpicture}
    \caption{Recall-QPS (1/s) tradeoff - up and to the right is better, 10-nearest neighbors,
        left: \textsf{SIFT-Hamming}, right: \textsf{Word2bits}. The following versions of \texttt{Annoy} 
    are shown in the plot:
    \texttt{A}, standard Annoy that uses Euclidean distance as its distance metric;
    \texttt{A (Ham.)}, Annoy with node splitting inspired by Bitsampling LSH and tuned to Hamming space; and
    \texttt{A (Eucl.)}, the run of \texttt{Annoy} on \texttt{SIFT} from Figure~\ref{plot:performance} (bottom left).}
    \label{plot:hamming}
\end{figure}
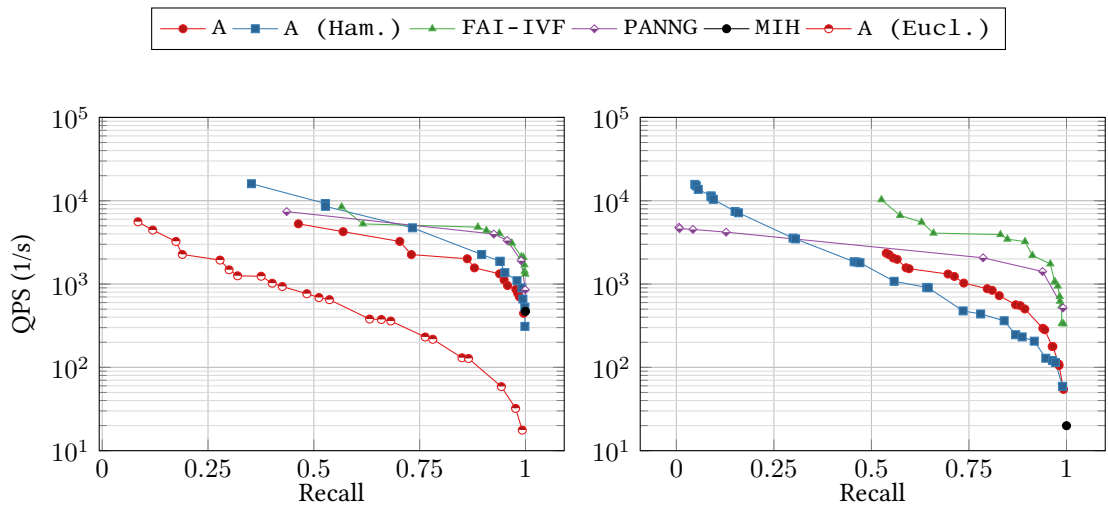

\noindent{\textbf{(Q3) Approximation.}} Figure~\ref{plot:approximation} relates achieved QPS to the (approximate) recall of an algorithm. The plots show 
results on the \textsf{GIST} dataset with 100-NN for recall with no
approximation and approximation factors of $1.01$ and $1.1$, respectively. Despite its high
dimensionality, all considered algorithms achieve close to perfect recall (left). For an
approximation factor of $1.01$, i.e., distances to true nearest neighbors are allowed to differ by $1\%$, all curves move to the right, as expected. Also, the relative difference between the performance of algorithms does not change. However, we see a clear difference between the candidate sets that are returned by algorithms at low recall. For example, the data point for \texttt{MRPT} around .5 recall on the left achieves roughly .6 recall
as a $1.01$ approximation, which means that roughly 10 new candidates are considered true approximate nearest neighbors. On the other hand, \texttt{HSNW}, \texttt{FAISS-IVF}, and \texttt{Annoy} improve by around 25 candidates being counted as approximate nearest neighbors. We see that allowing a slack of $10\%$ in the distance renders the queries too simple: almost all algorithms achieve near-perfect recall for all of their parameter choices. Interestingly, \texttt{Annoy} becomes the
second-fastest algorithm for $1.1$ approximation. This means that its candidates at very low recall values were a bit better than the ones obtained by its competitors.


\noindent{\textbf{(Q4) Embeddings.}} Figure~\ref{plot:hamming} shows a comparison between selected algorithms on the binary version of \textsf{SIFT} and a version of the Wikipedia dataset generated by \textsf{Word2Bits}, which is an embedding of \texttt{word2vec} vectors~\cite{Mikolov13} into binary vectors. The
performance plot for \texttt{Annoy} in the original Euclidean-space
version of \textsf{SIFT} is also shown. 

On \textsf{SIFT}, algorithms perform much faster 
in the embedded Hamming space version compared to
the original Euclidean-space version (see Figure~\ref{plot:performance}), which indicates that the queries are 
easier to answer in the embedded space. (Note here that the dimensionality is actually twice as large.)  
Multi-index
hashing \cite{mihalgo}, an exact algorithm for Hamming space, shows good performance on \textsf{SIFT} with around 460 QPS.

We created a Hamming space-aware version of \texttt{Annoy}, using
\texttt{popcount} for distance computations, and sampling single bits 
(as in Bitsampling LSH \cite{IndykM98}) instead of choosing hyperplanes. This version is two to three times faster on \textsf{SIFT} until high recall, where the Hamming space version and the Euclidean space version converge in running time.  
On the 800-dimensional \textsf{Word2Bits} dataset the opposite is true and the original version of \texttt{Annoy} is faster than the dedicated Hamming space approach. This means that the original data-dependent node splitting in \texttt{Annoy} adapts better to the query structure than the node splitting by data-independent Bitsampling for this dataset. The dataset seems to be hard in general: \texttt{MIH} achieves only around 20 QPS on \textsf{Word2Bits}.
We remark that setting the parameters for \texttt{MIH} correctly is crucial;
even though the recall will always be 1, different parameter settings can give
wildly different QPS values.

The embedding into Hamming space does have some consistent benefits that we do
not show here. Hamming space-aware algorithms should always have smaller index
sizes, for example, due to the compactness of bit vectors.

\subsection{Index build time remarks}

\begin{figure}[t!]
\begin{tikzpicture}
\selectcolormodel{gray}
  \begin{axis}[
    ybar,
    enlargelimits=0.15,
    legend style={at={(0.5,-0.2)},
      anchor=north,legend columns=-1},
    ylabel={Build time (s)},
    width = .9\textwidth,
    height= 5.5cm,
    symbolic x coords={
    FAISS-IVF,%
	ANNOY,%
    BALLTREE,%
	PyNNDescent,%
    PANNG,%
SW-Graph,%
HNSW(FAISS),%
RP-Forest,%
FLANN,%
KGraph,%
HNSW(NMSlib),%
},
    xtick=data,
    nodes near coords, 
    nodes near coords align={vertical},
    x tick label style={rotate=45,anchor=east},
    ]
    \addplot coordinates {
    (FAISS-IVF, 2.2)
    (ANNOY, 262)
	(BALLTREE, 287)
    (PyNNDescent, 765)
	(PANNG, 802)
    (SW-Graph, 1404)
    (HNSW(FAISS), 1700)
	(RP-Forest, 2667)
	(FLANN, 5032)
    (KGraph, 7189)
    (HNSW(NMSlib), 17000)
};
  \end{axis}
\end{tikzpicture}
\caption{Index build time in seconds for dataset \textsf{GLOVE}. The plot shows the minimum build time for an index that achieved recall of at least 0.9 for 10-NN.}
\label{fig:buildtime}
\end{figure}
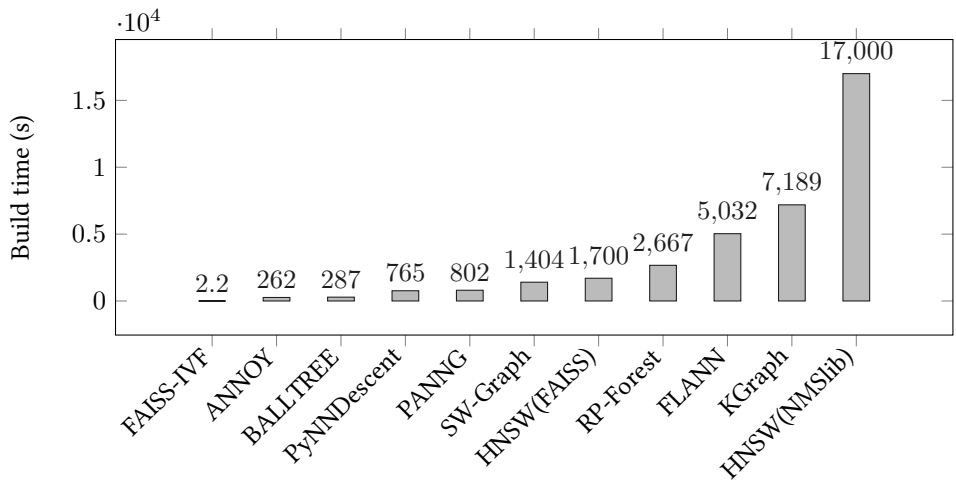

Figure~\ref{fig:buildtime} compares different implementations with respect to the time it takes to build the index. We see a huge difference in the index building time among implementations, ranging from \texttt{FAISS-IVF} (around 2 seconds to build the index) to \texttt{HNSW} (almost 5 hours). In general, building the nearest neighbor graph and building a tree data structure takes considerably longer than the inverted file approach taken by \texttt{FAISS-IVF}. Shorter build times make it much quicker to search for the best parameter choices for a dataset. Although all indexes achieve recall of at least 0.9, we did not normalize by the queries per second as in Figure~\ref{plot:index:size}. For example, \texttt{HNSW} also achieves its highest QPS with these indexes, but \texttt{FAISS} needs a larger index to achieve the performance from Figure~\ref{plot:performance} (which takes around 13 seconds to build). As an aside, building an HNSW index using the implementation provided in FAISS made it possible to build an index that achieved recall .9 in only 1\,700 seconds.

\subsection{Batched Queries}

We turn our focus to batched queries. In this setting, each algorithm is given the whole set of query points at once and has to return closest neighbors for each point. This allows for several optimizations: in a GPU setting, for example, copying query points to, and results from, the GPU's memory is expensive, and being able to copy everything at once drastically reduces this overhead.

The following experiments have been carried out on an Intel Xeon CPU E5-1650 v3 @ 3.50GHz with 6 physical cores, 15MB L3 Cache, 64 GB RAM, and equipped with an NVIDIA Titan XP GPU. 

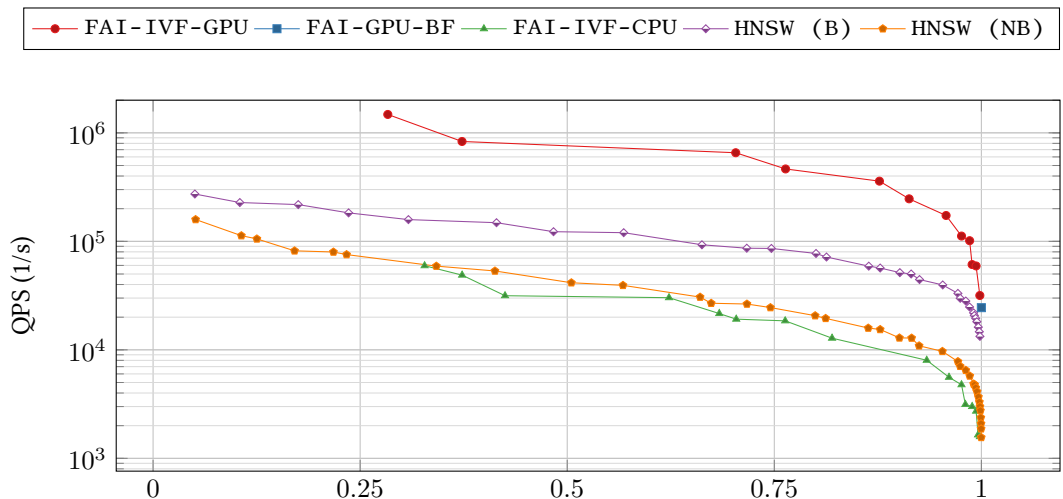
\begin{figure}
    \begin{tikzpicture}[every mark/.append style={mark size=1.5pt}]
        \begin{groupplot}[group style = {group size = 1 by 1, group name = hamming, vertical sep=2.1cm}, grid = both, grid style={line width=.1pt, draw=gray!30},
    major grid style={line width=.2pt,draw=gray!50},height=6.5cm, width=\textwidth,
              ymin = 0, ymax = 2000000, xtick = {0, 0.25, 0.5, 0.75, 1}, max space between ticks=20]

        \nextgroupplot[
            ylabel={QPS (1/s)},
            ylabel style ={overlay},
            ymode = log,
            yticklabel style={/pgf/number format/fixed,
                              /pgf/number format/precision=3},
            legend style = { at = { (1, 1.25 ) }},
            legend columns = 5
            ]
\addplot coordinates {
                                                (0.28338, 1477752.1756)
                                    (0.37301, 832698.828668)
                                    (0.70343, 655646.845495)
                                    (0.76362, 464856.142217)
                                    (0.87695, 358794.183062)
                                    (0.91279, 246504.810433)
                                    (0.9574, 173090.897086)
                                    (0.97605, 111871.674299)
                                    (0.98584, 101380.747272)
                                    (0.98882, 61130.3188195)
                                    (0.99363, 59330.6617139)
                                    (0.99816, 31690.2109732)
                            };
            \addlegendentry{ FAI-IVF-GPU };
                                \addplot coordinates {
                                                (1.0, 24471.5942374)
                            };
            \addlegendentry{ FAI-GPU-BF };
                                \addplot coordinates {
                                                (0.32781, 59623.916072)
                                    (0.37311, 48826.2185007)
                                    (0.42492, 31545.750056)
                                    (0.6229, 30246.5702847)
                                    (0.68356, 21606.3023602)
                                    (0.70405, 19177.8449304)
                                    (0.76315, 18507.0066932)
                                    (0.81975, 12807.7546952)
                                    (0.93405, 8017.92183418)
                                    (0.96095, 5585.82448731)
                                    (0.97603, 4769.34920272)
                                    (0.9807, 3139.09548363)
                                    (0.98887, 3003.50261506)
                                    (0.99366, 2722.67465935)
                                    (0.99603, 1644.80200631)
                                    (0.9982, 1585.14279348)
                            };
                        \addlegendentry{ FAI-IVF-CPU} ;
                                \addplot coordinates {
                                                (0.05046, 273008.012601)
                                    (0.10477, 227858.427271)
                                    (0.17526, 218131.816125)
                                    (0.23606, 182972.010889)
                                    (0.30827, 158641.995855)
                                    (0.41463, 148544.553053)
                                    (0.48345, 122737.006593)
                                    (0.5682, 120190.961974)
                                    (0.66264, 92958.4531983)
                                    (0.7168, 86425.9662976)
                                    (0.74649, 86023.0649802)
                                    (0.80038, 77320.3455383)
                                    (0.81302, 71716.5060828)
                                    (0.86408, 59079.6968758)
                                    (0.87785, 56784.2975855)
                                    (0.9016, 51493.3587712)
                                    (0.91524, 50023.2446918)
                                    (0.92533, 44443.0504753)
                                    (0.95336, 39641.0074173)
                                    (0.97168, 33124.5286178)
                                    (0.97446, 29879.1953576)
                                    (0.98123, 28174.7270729)
                                    (0.98574, 24842.2536463)
                                    (0.99058, 21934.2470023)
                                    (0.99152, 20741.9378608)
                                    (0.99311, 19539.8726134)
                                    (0.99453, 18159.8566367)
                                    (0.99632, 16204.1967141)
                                    (0.99756, 14676.5086158)
                                    (0.99827, 13307.764648)
                            };
            \addlegendentry{ HNSW (B)};
                                \addplot coordinates {
                                                (0.05133, 158952.219258)
                                    (0.10659, 112803.173536)
                                    (0.12532, 105351.964092)
                                    (0.1708, 81784.8632923)
                                    (0.21775, 79776.6264199)
                                    (0.23362, 75515.6259565)
                                    (0.342, 59100.9250583)
                                    (0.41276, 53335.571383)
                                    (0.50499, 41538.7845488)
                                    (0.56733, 39311.7474933)
                                    (0.66044, 30633.5141704)
                                    (0.67382, 26877.3121047)
                                    (0.71696, 26447.7351433)
                                    (0.74534, 24524.4794334)
                                    (0.79935, 20647.4276519)
                                    (0.81199, 19506.2586182)
                                    (0.86343, 15883.2254774)
                                    (0.87784, 15407.7052044)
                                    (0.90107, 12888.1577047)
                                    (0.9158, 12847.4762236)
                                    (0.92483, 10894.777172)
                                    (0.95296, 9685.39078319)
                                    (0.9717, 7802.39241143)
                                    (0.97441, 7056.67950259)
                                    (0.98118, 6485.38289205)
                                    (0.98597, 5748.56628621)
                                    (0.99068, 4849.03248928)
                                    (0.99181, 4763.14767413)
                                    (0.99323, 4471.76208027)
                                    (0.99476, 4122.57044966)
                                    (0.99644, 3683.1016458)
                                    (0.9976, 3306.24417074)
                                    (0.99829, 3006.18617379)
                                    (0.99885, 2757.67282958)
                                    (0.99932, 2372.69711192)
                                    (0.99955, 2087.85488658)
                                    (0.9997, 1871.84068889)
                                    (0.99984, 1558.72764569)
                            };
            \addlegendentry{ HNSW (NB)};
    \end{groupplot}
\end{tikzpicture}
    \caption{Recall-QPS (1/s) tradeoff - up and to the right is better. Algorithms running 
        batched queries on \textsf{SIFT} with $10$-NN. The plot shows a comparison between FAISS' IVF index 
    	running on a CPU and a GPU, FAISS' brute force index on the GPU, and HNSW from NMSlib running 
    in batched (B) and non-batched mode (NB).}
    \label{plot:batch}
\end{figure}

Figure~\ref{plot:batch} reports on our results with regard to algorithms in batch mode. \texttt{FAISS}' inverted file index on the GPU is by far the fastest index, answering around 655\,000 queries per second for .7 recall, and 61\,000 queries per second for recall .99. It is around 20 to 30 times faster than the respective data structure running on the CPU. Comparing \texttt{HNSW}'s performance with batched queries against non-batched queries shows a speedup by a factor of roughly 3 at .5 recall, and a factor of nearly
5 at recall .99 in favor of batched queries. Attention should also be put on the fact that the brute force variant of \texttt{FAISS} on the GPU answers around 24\,000 queries per second.  

\subsection{Summary}

\noindent\textbf{Which method to choose?} From the evaluation, we see that graph-based algorithms provide by far the best performance on most of the datasets. \texttt{HNSW} is often the fastest algorithm, but \texttt{PANNG} is more robust if there is no global structure in the dataset. The downside of graph-based approaches is the high preprocessing time needed to build their data structures. This could mean that they might not be the preferred choice if the dataset changes regularly. When it comes to small and quick-to-build index data structures, \texttt{FAISS}' inverted file index provides a suitable choice that still gives good performance in answering queries.

\medskip

\noindent\textbf{How well do these results generalize?} In our experiments, we observed that, for the standard datasets under consideration, algorithms usually agree on 
\begin{itemize}
\item [(i)] the order in how well they perform on datasets, i.e., if algorithm \texttt{A} answers queries on dataset \textsf{X} faster than on dataset \textsf{Y}, then so will algorithm \texttt{B}; and 
\item [(ii)] their relative order to each other, i.e., if algorithm \texttt{A} is faster than \texttt{B} on dataset \textsf{X}, this will most likely be the order for dataset \textsf{Y}.
\end{itemize}
There exist exceptions from this rule, e.g., for the dataset \textsf{Rand-Euclidean} described above.

\medskip

\noindent\textbf{How robust are parameter choices?} With very few exceptions (see Table~\ref{tab:algorithms}), users often have to set many parameters themselves. Of course, our framework allows them to choose the best parameter choice by exploring the interactive plots that contain the parameter choices that achieve certain quality guarantees. 

In general, the \emph{build parameters} can be used to estimate the size of the index\footnote{As an
example, the developers of \texttt{FAISS} provide a detailed description of the space usage of their indexes at \url{https://github.com/facebookresearch/faiss/wiki/Faiss-indexes}.},
while the \emph{query parameters} suggest the amount of effort that is put into searching the index. 

We will concentrate for a moment on Figure~\ref{plot:parameters}. This figure presents a scatter plot of selected algorithms for \textsf{GLOVE} on 10-NN, cf. the Pareto curve in Figure~\ref{plot:performance} (in the top left). Each algorithm has a very distinctive parameter space plot. 

For \texttt{HNSW}, almost all data points lie on the Pareto curve. This means that the different build parameters blend seamlessly into each other.
For \texttt{Annoy}, we see that data points are grouped into clusters of three points each, which represent exactly the three different index
choices that are built by the algorithm.
For low recall, there is a big performance penalty for choosing a too large index; at high recall, the different build parameters blend almost into each other.
For \texttt{SW-Graph}, we see two groups of data points, representing two different index choices.
We see that with the index choice to the left, only very low recall is achieved on the dataset.
Extrapolating from the curve, choosing query parameters that would explore a large part of the index will probably lead to low QPS. No clear picture is visible for \texttt{FAISS-IVF} from the plot. This is chiefly because we test many different build parameters -- recall that the index building time is very low. Each build parameter has its very own curve with respect to the different query parameters. 

As a rule of thumb, when aiming for high recall values, a larger index performs
better than a smaller index and is more robust to the choice of query parameters.

\begin{figure}
    \input{plot-parameters}
\end{figure}

\section{Conclusion \& Further Work}\label{sec:conclusion}
We introduced \textsf{ANN-Benchmarks}, an automated benchmarking system for
approximate nearest-neighbor algorithms. We described the system and used it
to evaluate existing algorithms. Our evaluation showed that well-enginereed
solutions for Euclidean and Cosine distance exist, and many techniques allow
for fast nearest-neighbor search algorithms. At the moment, graph-based
approaches such as \texttt{HNSW} or \texttt{KGraph} outperform the other
approaches for very high recalls, except for on very few datasets.
Index building for graph-based approaches takes a long time for
datasets with difficult queries.

In future, we aim to add support for other metrics and quality measures, such
as positional errors \cite{ZezulaSAR98}. Preliminary support
exists for set similarity under Jaccard distance, but algorithm implementations
are missing. Additionally, similarity joins 
are an interesting variation of the problem worth benchmarking \cite{ChristianiPS18}. We remark that the data we store with each algorithm run allows for more analysis beyond looking at average query times. One could, for example, already look at the variance of running times between algorithms, which could yield insights when comparing different approaches.
We also intend
to simplify and further automate the process of re-running benchmarks
when new versions of algorithm implementations appear. 

As a general direction for future work, we remark that none of the most performant implementations are easy to use. From a user perspective, the internal parameters of the data structure would ideally be invisible; an algorithm should be able to tune itself for the dataset at hand, given just a handful of quality-related parameters (such as the desired recall or the index size).

In the future, we plan to include a benchmarking mode for investigating this.
A new tuning step will be added to the framework, letting implementations
examine a small part of the dataset and to tune themselves for some given
quality parameters before training begins.
Implementors of algorithms would be able to test their auto-tuning techniques
easily with such a benchmarking mode.

Another general direction for future work is to get a better understanding which properties of a dataset make it easy or difficult for a specific algorithm. As shown in the evaluation, for many of the real-world datasets we get a homogeneous picture of how well algorithms perform against each other. On the other hand, we have given an example for a random dataset where implementations behave very differently. Which properties of a dataset make it simple or difficult
for a specific algorithmic approach? For example, for graph-based algorithms there has been very little research except \cite{Laarhoven18} on the theoretical guarantees that they achieve. 

Finally, answering batched queries, in particular on the GPU, is an interesting area for future work. There exist both novel LSH-based implementations \cite{WangSWR18} of nearest-neighbor algorithms and ideas on how to parallelize queries beyond running each query individually \cite{ChristianiPS18}. In particular, for a batch of queries an algorithm should exploit that individual queries might be close to each other.

\smallskip

\noindent\textbf{Acknowledgements:} We thank the anonymous reviewers for their careful comments
that allowed us to improve the paper. The first and third authors thank all members 
of the algorithm group at the IT University of Copenhagen for fruitful discussions.
In particular, we thank Rasmus Pagh for the suggestion of the random dataset.
This work was supported by a GPU donation from NVIDIA.

\bibliographystyle{splncs03}
\bibliography{lit}

\begin{thebibliography}{10}
\providecommand{\url}[1]{\texttt{#1}}
\providecommand{\urlprefix}{URL }

\bibitem{mrpt}
{MRPT} - fast nearest neighbor search with random projection,
  \url{https://github.com/teemupitkanen/mrpt}

\bibitem{ngt}
{NGT: PANNG}, \url{https://github.com/yahoojapan/NGT}

\bibitem{AhleAP17}
Ahle, T.D., Aum{\"{u}}ller, M., Pagh, R.: Parameter-free locality sensitive
  hashing for spherical range reporting. In: {SODA}'17. pp. 239--256

\bibitem{AlmanW15}
Alman, J., Williams, R.: Probabilistic polynomials and hamming nearest
  neighbors. In: FOCS'15. pp. 136--150

\bibitem{falconn}
Andoni, A., Indyk, P., Laarhoven, T., Razenshteyn, I.P., Schmidt, L.: Practical
  and optimal {LSH} for angular distance. In: {NIPS}'15. pp. 1225--1233.
  \url{https://falconn-lib.org/}

\bibitem{kdtree}
Bentley, J.L.: Multidimensional binary search trees used for associative
  searching. Commun. ACM  18(9),  509--517 (1975)

\bibitem{annoy}
Bernhardsson, E.: Annoy, \url{https://github.com/spotify/annoy}

\bibitem{nmslib}
Boytsov, L., Naidan, B.: Engineering efficient and effective non-metric space
  library. In: {SISAP}'13. pp. 280--293

\bibitem{BoytsovNMN16}
Boytsov, L., Novak, D., Malkov, Y., Nyberg, E.: Off the beaten path: Let's
  replace term-based retrieval with k-nn search. In: {CIKM}'16. pp. 1099--1108

\bibitem{ChristianiPS18}
Christiani, T., Pagh, R., Sivertsen, J.: Scalable and robust set similarity
  join. In: {ICDE}'2018 (2018)

\bibitem{CiacciaPZ97}
Ciaccia, P., Patella, M., Zezula, P.: M-tree: An efficient access method for
  similarity search in metric spaces. In: {VLDB}'97. pp. 426--435 (1997)

\bibitem{mlpack2013}
Curtin, R.R., Cline, J.R., Slagle, N.P., March, W.B., Ram, P., Mehta, N.A.,
  Gray, A.G.: {MLPACK}: A scalable {C++} machine learning library. Journal of
  Machine Learning Research  14,  801--805 (2013)

\bibitem{kgraph}
Dong, W.: {KGraph}, \url{https://github.com/aaalgo/kgraph}

\bibitem{mplsh}
Dong, W., Wang, Z., Josephson, W., Charikar, M., Li, K.: Modeling {LSH} for
  performance tuning. In: {CIKM}'08. pp. 669--678. ACM,
  \url{http://lshkit.sourceforge.net/}

\bibitem{edel2014automatic}
Edel, M., Soni, A., Curtin, R.R.: An automatic benchmarking system. In: NIPS
  2014 Workshop on Software Engineering for Machine Learning (2014)

\bibitem{HeoLHCY15}
Heo, J.P., Lee, Y., He, J., Chang, S.F., Yoon, S.E.: Spherical hashing: Binary
  code embedding with hyperspheres. {IEEE TPAMI}  37(11),  2304--2316 (2015)

\bibitem{Hyvonen2016}
Hyv{\"o}nen, V., Pitk{\"a}nen, T., Tasoulis, S., J{\"a}{\"a}saari, E.,
  Tuomainen, R., Wang, L., Corander, J., Roos, T.: Fast nearest neighbor search
  through sparse random projections and voting. In: Big Data (Big Data), 2016
  IEEE International Conference on. pp. 881--888. IEEE (2016)

\bibitem{IndykM98}
Indyk, P., Motwani, R.: Approximate nearest neighbors: {T}owards removing the
  curse of dimensionality. In: STOC'98. pp. 604--613

\bibitem{Iwasaki16}
Iwasaki, M.: Pruned bi-directed k-nearest neighbor graph for proximity search.
  In: {SISAP} 2016. pp. 20--33 (2016),
  \url{https://doi.org/10.1007/978-3-319-46759-7_2}

\bibitem{faiss}
Johnson, J., Douze, M., J{\'{e}}gou, H.: Billion-scale similarity search with
  gpus. CoRR  abs/1702.08734 (2017)

\bibitem{JohnsonL86}
Johnson, W.B., Lindenstrauss, J., Schechtman, G.: Extensions of lipschitz maps
  into banach spaces. Israel Journal of Mathematics  54(2),  129--138 (1986)

\bibitem{KriegelSZ17}
Kriegel, H., Schubert, E., Zimek, A.: The (black) art of runtime evaluation:
  Are we comparing algorithms or implementations? Knowl. Inf. Syst.  52(2),
  341--378 (2017)

\bibitem{Laarhoven18}
Laarhoven, T.: {Graph-Based Time-Space Trade-Offs for Approximate Near
  Neighbors}. In: Speckmann, B., T{\'o}th, C.D. (eds.) 34th International
  Symposium on Computational Geometry (SoCG 2018). Leibniz International
  Proceedings in Informatics (LIPIcs), vol.~99, pp. 57:1--57:14. Schloss
  Dagstuhl--Leibniz-Zentrum fuer Informatik, Dagstuhl, Germany (2018),
  \url{http://drops.dagstuhl.de/opus/volltexte/2018/8770}

\bibitem{Lam18}
Lam, M.: Word2bits - quantized word vectors. CoRR  abs/1803.05651 (2018),
  \url{http://arxiv.org/abs/1803.05651}

\bibitem{Lecun98}
LeCun, Y., Bottou, L., Bengio, Y., Haffner, P.: Gradient-based learning applied
  to document recognition. Proceedings of the IEEE  86(11),  2278--2324 (1998)

\bibitem{LiZSWZL16}
Li, W., Zhang, Y., Sun, Y., Wang, W., Zhang, W., Lin, X.: Approximate nearest
  neighbor search on high dimensional data - experiments, analyses, and
  improvement (v1.0). CoRR  abs/1610.02455 (2016),
  \url{http://arxiv.org/abs/1610.02455}

\bibitem{rpforest}
{Lyst Engineering}: Rpforest, \url{https://github.com/lyst/rpforest}

\bibitem{hnsw}
{Malkov}, Y.A., {Yashunin}, D.A.: {Efficient and robust approximate nearest
  neighbor search using Hierarchical Navigable Small World graphs}. ArXiv
  e-prints  (Mar 2016)

\bibitem{swgraph}
Malkov, Y., Ponomarenko, A., Logvinov, A., Krylov, V.: Approximate nearest
  neighbor algorithm based on navigable small world graphs. Inf. Syst.  45,
  61--68 (2014)

\bibitem{pynndescent}
McInnes, L.: {PyNNDescent}, \url{https://github.com/lmcinnes/pynndescent}

\bibitem{Mikolov13}
Mikolov, T., Sutskever, I., Chen, K., Corrado, G.S., Dean, J.: Distributed
  representations of words and phrases and their compositionality. In:
  {NIPS}'13. pp. 3111--3119

\bibitem{flann}
Muja, M., Lowe, D.G.: Fast approximate nearest neighbors with automatic
  algorithm configuration. In: {VISSAPP}'09. pp. 331--340. INSTICC Press

\bibitem{mihalgo}
Norouzi, M., Punjani, A., Fleet, D.J.: Fast search in hamming space with
  multi-index hashing. In: {CVPR}'12. pp. 3108--3115. IEEE

\bibitem{Pham17}
Pham, N.: Hybrid {LSH:} faster near neighbors reporting in high-dimensional
  space. In: {EDBT}'17. pp. 454--457

\bibitem{openml2013}
Van~Rijn, J.N., Bischl, B., Torgo, L., Gao, B., Umaashankar, V., Fischer, S.,
  Winter, P., Wiswedel, B., Berthold, M.R., Vanschoren, J.: Openml: A
  collaborative science platform. In: {ECML PKDD}. pp. 645--649. Springer
  (2013)

\bibitem{wang}
Wang, J., Shen, H.T., Song, J., Ji, J.: Hashing for similarity search: {A}
  survey. CoRR  abs/1408.2927 (2014), \url{http://arxiv.org/abs/1408.2927}

\bibitem{WangSWR18}
Wang, Y., Shrivastava, A., Wang, J., Ryu, J.: Randomized algorithms accelerated
  over {CPU-GPU} for ultra-high dimensional similarity search. In: {SIGMOD}.
  pp. 889--903 (2018), \url{http://doi.acm.org/10.1145/3183713.3196925}

\bibitem{Williams05}
Williams, R.: A new algorithm for optimal 2-constraint satisfaction and its
  implications. Theor. Comput. Sci.  348(2-3),  357--365 (2005)

\bibitem{ZezulaSAR98}
Zezula, P., Savino, P., Amato, G., Rabitti, F.: Approximate similarity
  retrieval with {M}-{T}rees. {VLDB} J.  7(4),  275--293 (1998)

\end{thebibliography}

\end{document}